\documentclass[
showpacs,
preprint,
aps,
prl,
superscriptaddress
]{revtex4-1}

\usepackage{graphicx}

\newcommand{\gtwotz}{g^{(2)}(\tau=0)}

\newcommand{\gtwoz}{g^{(2)}(0)}

\begin{document}

\title{
Distinguishing photon and polariton lasing from GaAs microcavities by spectral and temporal analysis of the two-threshold behavior}

\author{J.-S.~Tempel}
\author{F.~Veit}
\author{M.~A{\ss}mann}
\author{L.~E.~Kreilkamp}
\affiliation{Experimentelle Physik 2, Technische Universit\"{a}t Dortmund, D-44221 Dortmund, Germany}
\author{A.~Rahimi-Iman}
\author{A.~L{\"o}ffler}
\author{S.~H{\"o}fling}
\author{S.~Reitzenstein}
\author{L.~Worschech}
\author{A.~Forchel}
\affiliation{Technische Physik, Physikalisches Institut, Wilhelm Conrad R\"{o}ntgen Research Center for Complex Material Systems, Universit\"{a}t W\"{u}rzburg, D-97074 W\"{u}rzburg, Germany}  
\author{M.~Bayer}
\affiliation{Experimentelle Physik 2, Technische Universit\"{a}t Dortmund, D-44221 Dortmund, Germany}         

\date{\today}

\begin{abstract}
We compare polariton lasing with photon lasing of a planar GaAs/GaAlAs microcavity with zero detuning between the bare cavity mode and the bare exciton mode. For the emission from the lower energy-momentum dispersion branch we find a two-threshold behavior of the ground state in the input-output curve where each transition is accompanied by characteristic changes of the in-plane mode dispersion. In particular, we show that the thresholds are unambiguously evidenced in the photon statistics of the emission based on the second-order correlation function. Moreover, the distinct two-threshold behavior is confirmed in the evolution of the emission pulse duration.
\end{abstract}



\maketitle
Exciton-polariton lasing in semiconductor microcavities is an auspicious candidate for coherent light generation without the necessity to reach electronic population inversion \cite{Deng2003,Bajoni2008a}. Cavity polaritons are exciton-photon mixed quasi-particles resulting from the strong coupling (SC) of an optical cavity mode and a quantum-well exciton state. Being bosons, polaritons obey Bose-Einstein statistics and are supposed to undergo Bose-Einstein condensation (BEC). Due to their small effective mass, polariton condensates and polariton lasers may be realized at technologically relevant temperatures (even up to room temperature in wide band gap materials \cite{Malpuech2002}). But unlike BEC in atomic systems, where the phase transition occurs in thermal equilibrium \cite{Anderson1995,Davis1995}, polaritons are subject to strong dephasing and decay. Therefore, the polariton system requires external pumping to maintain condensate population. Despite this dissipative and non-equilibrium character, degenerate polariton systems show several textbook features of BEC, including the build-up of a macroscopic population of the lower polariton (LP) ground state \cite{Kasprzak2006}, spontaneous coherence \cite{Richard2005}, spatial condensation \cite{Balili2007}, superfluidity \cite{Amo2009a}, and vorticity \cite{Lagoudakis2008}.

However, it is hard to distinguish between coherent light emission caused by a condensed polariton system and coherent emission from a conventional photon laser. To demonstrate true polariton lasing one has to ensure the system to stay in the SC regime. Thus, the effective exciton oscillator strength and hence the Rabi splitting have to be sufficiently large as to prevent the system to pass on to the weak coupling regime \cite{Houdre1995,Butte2002}. A regime called stimulated polariton photoluminescence has been reported on a CdTe based system \cite{Dang1998}. In III$-\!$V materials, two different regimes of polariton and photon lasing have been demonstrated on the 3D-confined states of a micropillar cavity \cite{Bajoni2008a}. Also, it has been claimed that two different transitions have been observed in a planar GaAs microcavity, but only when stress is applied to the sample \cite{Balili2009,Nelsen2009}.

Nevertheless, a thorough examination of the different emission regimes of an unstressed planar III-V microcavity has not been reported so far. The observation of two distinct thresholds---the first reflecting the transition to a coherent condensate, the second manifesting saturation of the SC regime and the onset of photon lasing---is considered to be an unambiguous proof for polariton lasing in the excitation power range between the thresholds \cite{Dang1998,Bajoni2007}. In this letter, we present a detailed analysis of both nonlinearities observed on one position of a GaAs based microcavity. We give experimental results of power-dependent dispersion measurements as well as the corresponding photon quantum statistics obtained using a setup providing the appropriate time resolution. Additionally, we show that characteristic changes in the emission pulse duration accompany both thresholds.

The sample used for the experimental studies consists of a $\lambda/2$ cavity with three sets of $4$ GaAs/AlAs quantum wells. The cavity mirrors are composed of $16$ alternating GaAlAs/AlAs distributed Bragg reflector (DBR) layers on top and $20$ alternating layers at the bottom of the central cavity layer. The sample has a vacuum Rabi splitting of $\sim14\,$meV; detailed information on the sample design can be found in an earlier publication \cite{Utsunomiya2008}. All measurements were performed at zero energy detuning between cavity mode and exciton; the sample was kept in a helium-flow cryostat at a temperature of $10\,$K. Using a picosecond-pulsed Ti-Sapphire laser (repetition rate $75.39\,$MHz) the pump was focused to a spot approximately $30\,\mu$m in diameter on the sample at an angle of $45 ^\circ$ from normal incidence. For non-resonant excitation the pump laser was tuned to the first sideband minimum of the DBR structure at $\approx 744\,$nm. The emission of the sample was collected from the central area of the excitation spot using a microscope objective with a numerical aperture of $0.26$. To investigate the far field emission of the sample, the Fourier plane was imaged onto the entrance slit of a monochromator and detected with a nitrogen-cooled CCD camera.

Photon-correlation measurements were carried out using a streak camera with an additional horizontal deflection unit. The time resolution of $2\,$ps is adequate for the fast processes in semiconductor microcavities. Details of this technique were reported elsewhere \cite{Assmann2009,Assmann2010a}. A particular wavelength range of the emission was selected using an interference filter with a $1\,$nm wide transmission window, while a pinhole in the detection path guaranteed the exclusive collection of photons with in-plane momentum of $k_{||} = 0$.

\begin{figure}
\centering
\includegraphics[width=8.5cm]{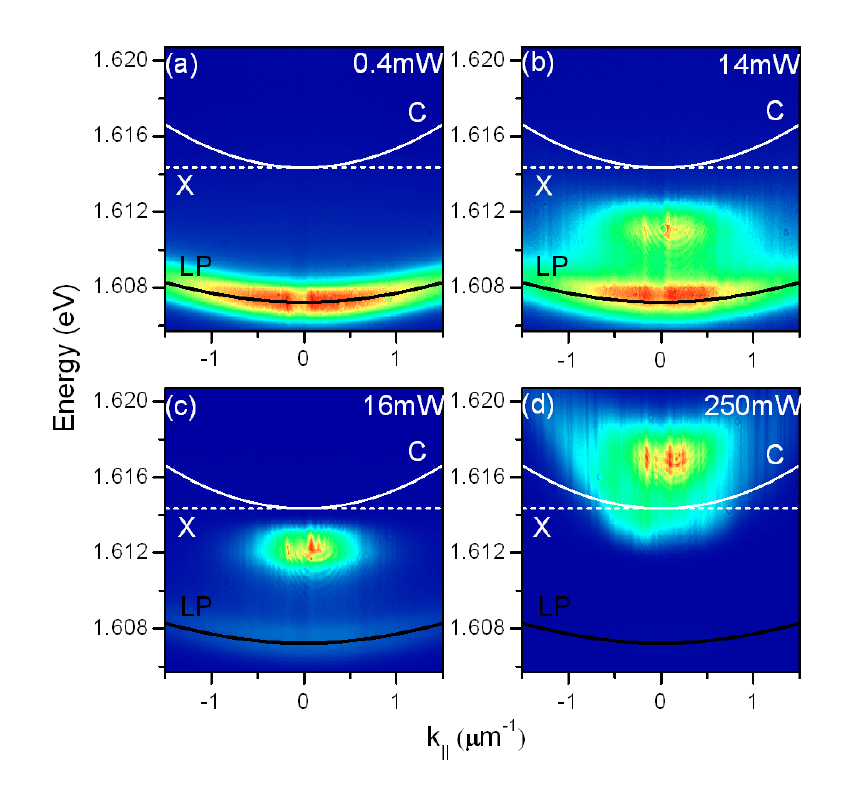}
\caption{\label{corr1}
(color online) Dispersion curves measured at different excitation powers. The lines give calculated LP (black), cavity photon (white, solid), and quantum-well exciton (white, dotted) mode dispersion, respectively. (a) At low excitation power ($0.4\,$mW) only the LP branch is populated. In the excitation power range from $14\,$mW (b) to $16\,$mW (c) the condensate starts to build-up. (d) Far above the condensation threshold power the SC is bleached and photon lasing from a blue shifted cavity-photon mode is observed.}
\end{figure}

Figure \ref{corr1} shows in-plane cavity mode dispersions at different excitation powers. At low pump powers (here $0.4\,$mW) only the LP is observed, as it is exclusively populated at low temperatures (panel (a)). There is evidence for the typical transition from a quadratic LP branch at low excitation power towards a blue shifted condensate at about $16\,$mW, see panels (b) and (c). At excitation powers well above this first threshold ($250\,$mW) the emission comes predominantly from a blue shifted cavity-photon mode---standard photon lasing sets in (panel (d)). In this regime the SC is finally bleached and the system remains weakly coupled. The blue shift with respect to the calculated bare cavity mode is usually attributed to changes in the material refractive indices with increasing carrier density \cite{Mohideen1994}.

\begin{figure}
\centering
\includegraphics[width=8.5cm]
{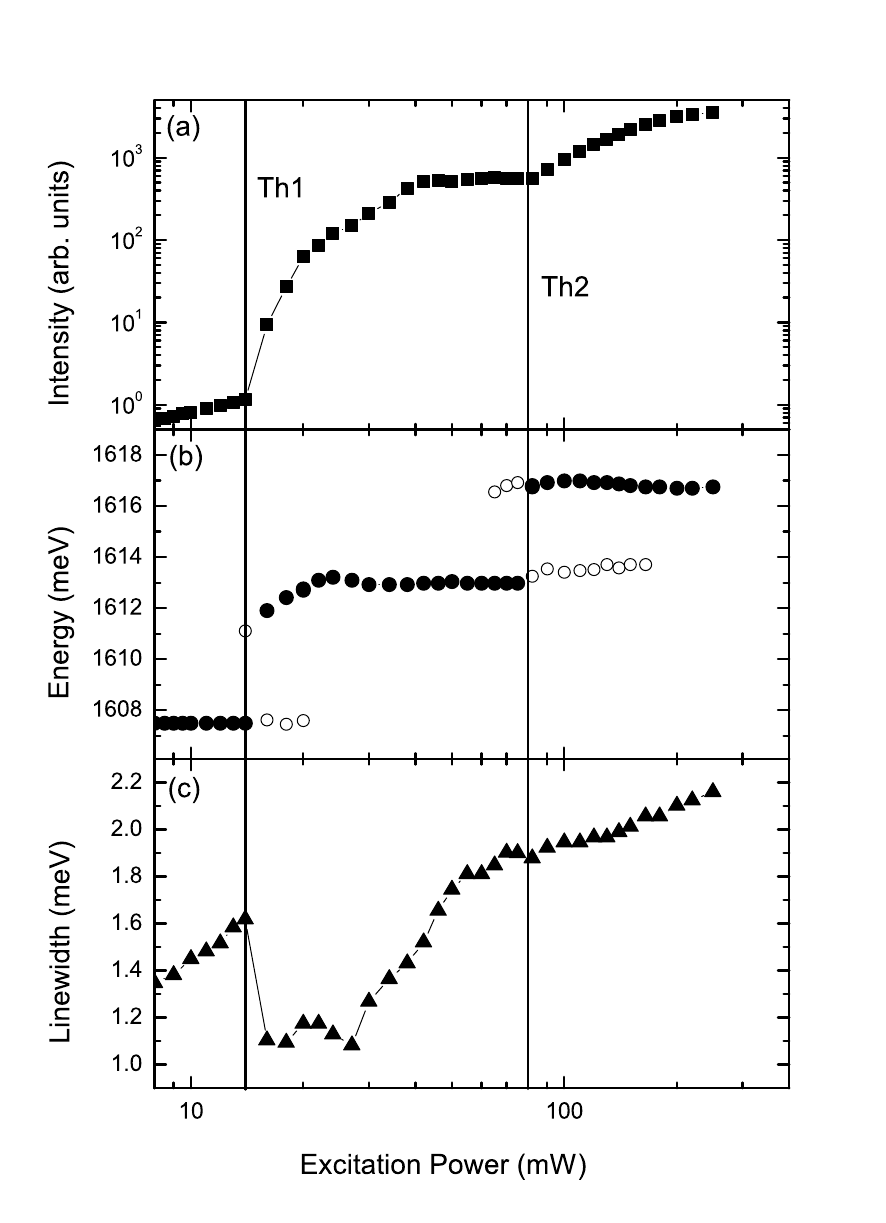}
\caption{\label{corr2}
Excitation power dependence of (a) the integrated intensity of the cavity emission with in-plane momenta $|k_{||}| \leq 0.13\,\mu$m$^{-1}$, (b) the emission energy at $k_{||} = 0$, and (c) the spectral linewidth (FWHM); the vertical lines indicate the two thresholds. The solid (open) symbols in panel (b) represent the energy of the mode with strongest (less) intensity; the pronounced jumps observed in the emission energy are attributed to non-resonant pulsed excitation.}
\end{figure}

In order to analyze the dispersion curves in more detail, the development of different parameters with increasing excitation power has been extracted, and the results are summarized in Fig.~\ref{corr2}. First of all, the integrated emission intensity as a function of the excitation power has been plotted, see Fig.~\ref{corr2}(a). Here, only the emission at zero in-plane momentum $(|k_{||}| \leq 0.13\,\mu$m$^{-1}$) has been taken into account. In the thermal regime the intensity increases linearly with pump power. With further increasing excitation power a strongly nonlinear behavior is observed, until a first saturation level is reached. At even higher pump powers a second nonlinear behavior is found. Both nonlinearities occur at the power levels where the changes in the dispersion curves are seen in Fig.~\ref{corr1}. A similar behavior is found in the development of the emission peak energy, as shown in Fig.~\ref{corr2}(b). Being rather constant in the thermal regime a clear jump of the emission energy of $\sim3.5\,$meV is found at the condensation threshold. In the condensed regime the emission energy first exhibits a further continuous blue shift and then remains at an almost constant value, until a second jump occurs at the same excitation power that was identified as the transition to photon lasing. The total blue shift as compared to the low density LP energy is then roughly $10\,$meV, the SC is thus definitively saturated.

Contrary to the two-threshold behavior in Figs.~\ref{corr2}(a) and (b), only one significant transition is found in the evolution of the spectral linewidth of the cavity emission. The full-width-at-half-maximum (FWHM) increases gradually in the thermal regime. A clear narrowing is observed at the transition to the condensate, here the linewidth is $\sim1.1\,$meV. Starting from about twice the threshold power, the linewidth increases continuously also throughout the regime of photon lasing. In the excitation power range between the two thresholds described above, the monotonous increase can be attributed to polariton-polariton interactions \cite{Porras2003}, as will be discussed below. The ongoing broadening of the linewidth at and above the transition to photon lasing can be explained by fluctuations of the refractive indices throughout an excitation pulse \cite{Wang2005}. Nevertheless, two distinct thresholds have been observed: one at $14\,$mW and another one at about $80\,$mW.

While these results are both clear indicators of the system changing its state at those excitation densities, a detailed characterization of the emission requires also studies of its coherence properties as well as its behavior in the time domain. The coherence properties have been determined in terms of the photon statistics of the cavity emission, as those are supposed to reflect the quantum statistical behavior of the polariton system. In particular, the second-order correlation function for equal-time events,
\begin{equation}
\gtwotz = \frac{\left\langle : \hat{n}^2 :\right\rangle}{\left\langle \hat{n} \right\rangle^2}, \label{corrfunc}
\end{equation}
was measured. In Eq.~(\ref{corrfunc}), $\hat{n}= \hat{a}^{\dagger}\hat{a}$ denotes the photon number operator and the double stops represent normal ordering of the photon creation $\hat{a}^{\dagger}$ and annihilation $\hat{a}$ operators \cite{Glauber1963}. Basically, one can distinguish between two different cases of photon statistics. Photons emitted from a thermal light source are expected to exhibit bunching behavior. As those photons are indistinguishable particles obeying the Bose-Einstein statistics, the corresponding second-order correlation function has a value of $g^{(2)} (0) = 2$. Photons emitted by coherent light sources are statistically independent particles, thus following the Poisson distribution. This is reflected in a photon correlation value of $\gtwoz =1$.

For microcavities with a vacuum Rabi splitting too small for the LPs to condense, a transition from $\gtwoz = 2$ to a value of $\gtwoz = 1$ with increasing excitation power was reported recently \cite{Assmann2009}. However, the power-dependent second-order correlation of a microcavity polariton system, with sufficiently large Rabi splitting to undergo BEC, is supposed to be more sophisticated. Investigations of the photon statistics presented so far have been performed with time resolutions of many tens to a few hundred picoseconds \cite{Deng2002,Kasprzak2008,Love2008}: A smooth transition towards a coherent state \cite{Deng2002} as well as increasing photon fluctuations above the condensation threshold \cite{Kasprzak2008} have been reported. By using the appropriate time resolution we resolve here the complete evolution of photon correlations over a broad range of excitation powers, which has not been reported so far.

\begin{figure}
\centering
\includegraphics[width=8.5cm]
{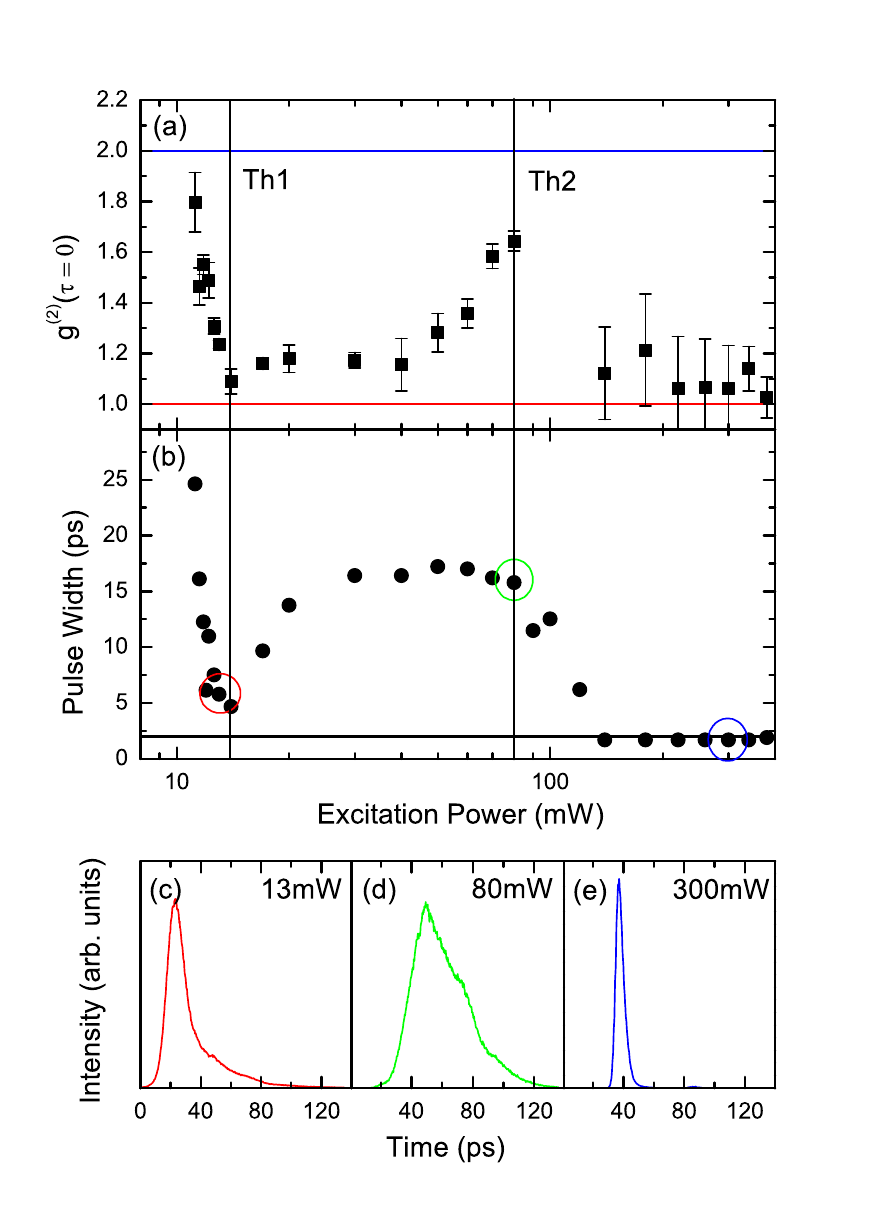}
\caption{\label{corr3}
(color online) (a) Power-dependent equal-time correlation function $\gtwotz$. The vertical lines represent the thresholds, the blue (red) line indicates the thermal (coherent) state. (b) Emission pulse DHM versus excitation power. The horizontal line indicates the limiting time resolution of the streak camera. (c)$-$(e) Temporal pulse profile at different excitation powers, also indicated by circles in (b).}
\end{figure}

Figure \ref{corr3}(a) shows the results of our photon-correlation measurements as a function of excitation power. At low pump powers the correlation function tends to a value of approximately $2$. Hence, the emission has thermal light characteristics. With increasing excitation power $\gtwoz$ decreases towards unity. This drop coincides with both the build-up of the condensate (Fig.~\ref{corr1}) and the first nonlinear increase of the integrated intensity [Fig.~\ref{corr2}(a)]. For a rather small range of excitation powers the system stays in a regime in which the emitted light exhibits almost full coherence, i.e.\ $\gtwoz$ values close to unity, before photon fluctuations increase again until $\gtwoz$ reaches a value of about $1.65$. This increase can be explained by scattering processes between the polaritons at $k_{||} = 0$ and polaritons with $k_{||} \neq 0$ which result in a depletion of the condensate ground state despite of the increasing pumping \cite{Schwendimann2008,Sarchi2008}. This is consistent with the saturation behavior seen in the input-output curve. With a further increase of the pump power the second-order correlation function approaches $\gtwoz = 1$ again, reflecting the breakdown of SC and the transition to conventional photon lasing.

The two-threshold behavior seen in the input-output curve and in the second-order correlation is confirmed by the evolution of the cavity's emission temporal pulse width with increasing excitation power, as shown in Fig.~\ref{corr3}(b). Because of the strong asymmetry of some pulses and non-exponential decay (Fig.~\ref{corr3}, bottom row), the pulse duration-at-half-maximum (DHM) was used to determine the pulse width. At low pump powers the pulses are comparatively long ($25\,$ps and more) and the LP branch gets slowly populated through spontaneous scattering processes. With increasing power the pulses are getting shorter because of parametric scattering into the condensate ground state. The range where the pulse duration is as short as $5\,$ps is relatively small. At those excitation powers the emitted light shows almost full coherence, the system is in the condensed state. Starting from $16\,$mW polariton-polariton scattering processes set in and cause a depletion of the $k_{||} = 0$ state, the pulse duration is getting longer again and reaches a saturation level at about $17\,$ps.
While stimulated scattering channels down to the condensate ground state remain on for a longer time window, the pulse DHM of $17\,$ps is supposed to be limited by the time after which all free carriers have relaxed down to the LP branch.
In this regime above the first threshold, the occupation of the ground state is supposed to stay at a certain level above unity but does not increase further. Considering that photon correlations concurrently increase, we note that the DHM is an adequate additional measure to fully characterize the system and the contributing scattering processes. Furthermore, simultaneously with the second threshold in the integrated intensity and the decrease of second-order correlation to a value of $1$, the pulses are shortening again. The shortest pulse width of $2\,$ps is given by the bare cavity lifetime. Thus, both thresholds are accompanied by distinct changes in the temporal pulse width.

To conclude, the two-threshold behavior in the emission of a planar GaAs microcavity under non-resonant pumping has been demonstrated and analyzed in detail. The comparison of five different quantities---measured on a single sample position without applying any stress---revealed a good overall accordance. At low powers the system is identified to be in a thermal regime of polariton photoluminescence, which is underlined by the long emission pulses. Simultaneously with the first nonlinearity in the input-output curve, the build-up of a lasing polariton condensate with short emission pulse duration was observed. With further increasing pump power, self-scattering processes of the polaritons led to an increase of second-order correlation to values significantly higher than unity, consistent with a saturation behavior in the emission pulse DHM. The second threshold at high excitation power went along with the breakdown of SC and the onset of conventional photon lasing, also reflected in extremely short pulse widths.
Thus, we summarize that the combination of photon statistics with a temporal and spectral analysis does give clear-cut criteria to characterize the quantum state of a microcavity polariton system.

The Dortmund group acknowledges support by the Deutsche Forschungsgemeinschaft through the research grant DFG 1549/15-1. The group at W\"urzburg University acknowledges support by the State of Bavaria.

%

\end{document}